\documentclass[twocolumn,aps,pra,superscriptaddress,a4paper,sort&compress,balancelastpage]{revtex4-2}
\usepackage[normalem]{ulem}
\usepackage{amsmath}
\usepackage{amssymb}
\usepackage{gensymb}
\usepackage{bm}
\usepackage{graphicx}
\usepackage{float}
\usepackage{subfigure}
\usepackage[dvipsnames]{xcolor}
\usepackage{placeins}
\usepackage{braket}
\usepackage{bbold}
\usepackage{ulem}
\usepackage[colorlinks, linkcolor=blue, citecolor=blue, urlcolor=blue, breaklinks=red]{hyperref}

\usepackage{bbm}

\usepackage{lipsum}
\usepackage{physics}
\usepackage{dsfont}

\definecolor{Mycolor1}{HTML}{44aa99}
\definecolor{Mycolor2}{HTML}{cc6677}
\usepackage{orcidlink}

\begin{document}

\title{Quantum correlations and metrological advantage among Unruh-DeWitt detectors in de Sitter spacetime}

\author{Samira Elghaayda~\!\!\orcidlink{0000-0002-6655-0465}}
\email{samira.elghaayda-etu@etu.univh2c.ma}
\affiliation{ Laboratory of High Energy Physics and Condensed Matter, Department of Physics,\\ Faculty of Sciences of Aïn Chock, Hassan II University,\\ P.O. Box 5366 Maarif, Casablanca 20100, Morocco.}
\author{Asad Ali\orcidlink{0000-0001-9243-417X}} \email{asal68826@hbku.edu.qa}
\affiliation{Qatar Center for Quantum Computing, College of Science and Engineering, Hamad Bin Khalifa University, Doha, Qatar}

\author{M. Y. Abd-Rabbou~\!\!\orcidlink{0000-0003-3197-4724}}
\email{m.elmalky@azhar.edu.eg}
\affiliation{School of Physics, University of Chinese Academy of Science,\\ Yuquan Road 19A, Beijing, 100049, China}
\affiliation{Mathematics Department, Faculty of Science, Al-Azhar University, \\ Nassr City 11884, Cairo, Egypt}

\author{Mostafa Mansour~\!\!\orcidlink{0000-0003-0821-0582}}
\email{mostafa.mansour.fsac@gmail.com}
\affiliation{ Laboratory of High Energy Physics and Condensed Matter, Department of Physics,\\ Faculty of Sciences of Aïn Chock, Hassan II University,\\ P.O. Box 5366 Maarif, Casablanca 20100, Morocco.}

\author{Saif Al-Kuwari\orcidlink{0000-0002-4402-7710}}
\email{smalkuwari@hbku.edu.qa}
\affiliation{Qatar Center for Quantum Computing, College of Science and Engineering, Hamad Bin Khalifa University, Doha, Qatar}

\begin{abstract}
A long-standing debate on Gibbons-Hawking (GH) decoherence centers on its obscure thermal nature. In this work, we investigate the robustness of quantum Fisher information (QFI) and local quantum uncertainty (LQU) in the presence of GH decoherence, using free-falling Unruh-DeWitt (UDW) detectors in de Sitter spacetime (dS-ST). The UDW detectors interact with a massless scalar field in dS-ST and are modeled as open quantum systems, with the field serving as the environment, described by a master equation that outlines their evolution. Our analysis investigates the roles of energy spacing, GH temperature, initial state preparation, and various de Sitter-invariant vacuum sectors on the optimization of QFI and LQU. We find that the optimal values of QFI and LQU depend on the selected de Sitter-invariant vacuum sector and increase with larger energy spacing. Our findings reveal that QFI exhibits resilience to GH decoherence, maintaining a pronounced local peak across a broader range of parameters. This robustness can be further enhanced through strategic initial state preparation and increased energy spacing, resulting in a higher maximum QFI value even under significant environmental decoherence. Our results underscore the critical role of GH thermality in governing QFI and LQU, offering valuable insights for advances in relativistic quantum metrology (RQM).
\end{abstract}


\maketitle


\section{Introduction}
Quantum information in curved spacetime (CST) plays a vital role in exploring the interplay between quantum physics and gravity \cite{bekenstein1973black, palmer2012localized, bekenstein2002quantum, aaElghaayda_2024, elghaayda2023entropy,wu2023would}. Although the abstract properties of quantum fields in CST, such as their entanglement \cite{hawking2001desitter, ball2006entanglement,ali2024quantum}, can be studied directly, operational approaches that involve observers and detectors have historically advanced theoretical insights \cite{birrell1984quantum, gibbons1977cosmological}. The Minkowski vacuum, for instance, exhibits long-range entanglement \cite{summers1987maximal}, which can be transferred to local inertial systems via standard quantum mechanical coupling mechanisms \cite{reznik2005violating}. Variants of this concept include CST \cite{martin2014entanglement}, thermal states \cite{braun2005entanglement}, and accelerating detectors \cite{massar2006einstein}. This work will focus on CST, particularly de sitter spacetime (dS-ST) \cite{bousso2002adventures, spradlin2002sitter, guth1998inflationary}, which is significant in cosmology. In CST, the definition of vacuum states is essential for analyzing vacuum fluctuations. In dS-ST, vacuum states fall into two categories: dS-invariant states and states that violate dS invariance \cite{allen1985vacuum}. The dS-invariant vacuum is widely regarded as the natural choice, as its role in dS-ST parallels that of the Minkowski vacuum in flat spacetime.

The Unruh-DeWitt (UDW) detector is a crucial tool for probing quantum fields within specific spacetime geometries \cite{dewitt1979quantum,birrell1984quantum,moustos2017non,juarez2019asymptotic}. This detector is a microscopic two-level system (qubit) that couples locally to fluctuating backgrounds. In dS-ST, a co-moving UDW detector detects radiation with a thermal spectrum defined by the temperature \( T_{GH} = \frac{H}{2\pi} \), where \( H \) is the Hubble parameter \cite{gibbons1977cosmological}. This phenomenon, known as the GH effect, demonstrates universality in various quantum fields. For a massless scalar field \cite{garbrecht2004unruh}, the infrared divergence of the propagator has negligible influence, leaving the detector's overall response largely unaffected. To explore the dynamic properties of quantum fields in CSTs \cite{birrell1984quantum}, the open quantum systems framework has gained prominence \cite{yu2011open, yu2008understanding}. This approach has proven valuable in examining the behavior of UDW detectors in diverse spacetime backgrounds \cite{benatti2004entanglement, hu2011entanglement, feng2015uncertainty,li2025does}. Within this framework, the UDW detector is modeled as a localized open quantum system, while the fluctuations of the background quantum field act as an environment that induces dissipation and dephasing.

In the framework of inflationary cosmology, quantum fluctuations are stretched during inflation, leading to anisotropies in the cosmic microwave background (CMB). The initial state during inflation is highly squeezed \cite{grishchuk1990squeezed}, and a Bell inequality test has been proposed to confirm the quantum mechanical origin of these primordial density fluctuations. Traditionally, these fluctuations are thought to arise from a Bunch-Davies vacuum (BDV) in the infinite past. However, this assumption has been questioned due to the inaccessibility of field modes below the Planck scale, $\Lambda$ \cite{brandenberger2013trans}. Consequently, inflation models incorporating non-BD initial conditions, including alternatives like dS-invariant vacua \cite{mottola1985particle, allen1985vacuum}, initial scalar field entanglement \cite{albrecht2014cosmological}, and correlated bubble universes \cite{kanno2015cosmological}, have attracted growing interest. A particularly intriguing feature of dS-ST is the existence of $\alpha$-vacua, an infinite class of dS-invariant vacuum states parameterized by a complex variable $\alpha$. These states, originally introduced decades ago \cite{mottola1985particle, allen1985vacuum}, recently received renewed attention despite ongoing debates about their ultraviolet behavior \cite{collins2004taming, goldstein2003note, einhorn2003interacting, einhorn2003squeezed, danielsson2002consistency}. As viable candidates for non-BD initial conditions, $\alpha$-vacua offer the potential for trans-Planckian modifications to the CMB \cite{martin2001trans, kaloper2003initial, goldstein2003initial}. Realizing this requires establishing a robust connection between the parameter $\alpha$ and the high-energy cutoff scale, $\Lambda$ \cite{danielsson2002inflation, danielsson2002note}.

In quantum estimation theory, a central challenge is estimating unobservable parameters in a labeled quantum system using measurement data \cite{gio2016,gio2011}. Enhancing the precision of these estimates and achieving the quantum accuracy limit are key objectives. Due to its practical relevance, the QFI has been explored from multiple perspectives, leading to significant progress. For instance, it has been used to establish a statistical generalization of the Heisenberg uncertainty principle \cite{luo2000quantum}. The measurement of QFI with finite precision was demonstrated in \cite{frowis201}, and its connections to quantum correlations (QCs) \cite{hyllus2012fisher,kim2018characterizing} and quantum coherence \cite{tan2018coherence} have been established. Recently, there has been a growing interest in integrating quantum metrology and quantum information within relativistic frameworks, particularly in exploring quantum fields in CSTs \cite{ahm,aaElghaayda_2024,elghaayda2024physically}. Furthermore, numerous studies have investigated the characteristics of QFI in non-inertial frames \cite{yao2014quantum,abd2023detraction}. By leveraging relativistic effects in quantum systems, it has shown that measurement precision can be greatly improved \cite{ahmad1}. This emerging discipline, referred to as RQM, seeks to identify the fundamental limits of measurement accuracy when quantum and relativistic effects are jointly considered. Significant relativistic phenomena explored in this field include the Unruh-Hawking effect \cite{aspachl,borim2n}, the expansion rate of Robertson-Walker universes \cite{wang20r}, and the properties of Schwarzschild spacetime \cite{bruscm}.  
\subsection{Contribution}
Working within the framework of detector-field interactions, we study a system of two UDW detectors in dS-ST. Assuming a weak interaction with a bath of scalar fields, the system behaves like an open quantum system, experiencing environmental decoherence due to the thermal nature of dS-invariant vacua, such as the GH effect. In this paper, we evaluate the LQU and QFI at the final equilibrium state of the system and analyze how factors such as the detector's energy spacing, GH decoherence, the choice of initial state, and the selection of $\alpha$-vacua influence LQU and QFI. Our findings highlight the robustness and enhanced sensitivity of LQU and QFI, illustrating their potential to enhance quantum estimation techniques in practical applications.

\subsection{Organization}
The remainder of the paper is structured as follows. Sec. \ref{sec1} introduces the model, in which two UDW detectors interact with a scalar field in a dS-ST background. The master equation in Lindblad form is solved to obtain the final equilibrium state. This section also examines fundamental concepts, including QFI and LQU for bipartite states. In Sec. \ref{sec2}, we explore the dynamics of UDW in non-BD vacua, which depend on the initial state preparation of the detectors, the GH decoherence effect, and different choices of superselection sectors of $\alpha$-vacua. Sec. \ref{sec3} presents our results, and Sect. \ref{sec4} concludes the paper.

\section{Preliminaries \label{sec1}}
This section presents the essential concepts for defining QFI and LQU for a localized UDW in dS-ST. We begin by describing the dynamics of a bipartite UDW governed by a Lindblad master equation. Next, we examine the definitions and properties of QFI and LQU for bipartite systems.

\subsection{Markovian dynamics of UDW detectors}
To proceed, we should first explore the Markovian evolution of an accelerating UDW detector interacting with a scalar field. The total Hamiltonian of the system, which includes the UDW detectors and the environment, is given by 
\begin{equation}
\mathcal{H}=\mathcal{H}_D+\mathcal{H}_\chi+\mathcal{H}_{int},
\end{equation}
Here, $\mathcal{H}_D$ represents the Hamiltonian of two independent detectors in the standard co-moving frame. The internal dynamics of each two-level detector is governed by a matrix $2 \times 2$, and $\mathcal{H}_D$ can be concisely expressed as \cite{elghaayda2023entropy,yu2011open,feng2015uncertainty,aaElghaayda_2024}.

\begin{equation}
	\mathcal{H}_D=\frac{\omega}{2}\left( {\hat{s}_{3}^{(a)}}\otimes\hat{s}_{0}^{(b)}+\hat{s}_{0}^{(a)}\otimes{\hat{s}_{3}^{(b)}}\right)\equiv \frac{\omega}{2}\Gamma_3,
\end{equation}
where the symmetrized bipartite operators, $\Gamma_i={\hat{s}_{i}^{(a)}}\otimes\hat{s}_{0}^{(b)}+\hat{s}_{0}^{(a)}\otimes{\hat{s}_{i}^{m}}$ are defined using Pauli matrices ${\hat{s}_{i}^{k}}(i=1,2,3)$, with superscript $m=\left\lbrace a,b\right\rbrace $ labeling distinct detectors, and $\omega$ is the energy spacing of the detectors. $\mathcal{H}_\chi$ is the Hamiltonian of free massless scalar fields $\chi(x)$ satisfying the Klein–Gordon equation $\Box \chi(x) = 0$ in dS-ST, with covariant d’Alembertian operator $\Box \equiv g^{\mu\nu} \nabla_\mu \nabla_\nu$ determined by the chosen coordinate system \cite{yu2011open,feng2018bell}. $\mathcal{H}_{int}$ describes the interaction between the system of two detectors and the scalar field, assumed to be in the form of electric dipole interaction
\begin{equation}
	\mathcal{H}_{int}=\lambda\left[  (\hat{s}_2^{(a)}\otimes\hat{s}_{0}^{(b)})\chi(t,\textbf{x}^{(a)})	+(\hat{s}_{0}^{(a)}\otimes\hat{s}_2^{(b)})\chi(t,\textbf{x}^{(b)})	\right],  
\end{equation}
where $\lambda$ is a small dimensionless coupling constant.
As we aim to study the dynamical evolution of detectors' density matrix $\eta_{ab}(t)=Tr_\chi[\eta_{tot}(t)]$, where $t$ is the proper time of detectors' world-line, we assume that the initial density matrix of the total system is separable, i.e., $\eta_{tot}(0)=\eta_{ab}(0)\otimes|0\rangle\langle 0|$, where $|0\rangle$ is the vacuum state of field $\chi(x)$ and $\eta_{ab}(0)$ is the system's initial density matrix. The complete system, including UDW detectors and bath, constitutes a closed system. Therefore, the total density matrix evolves according to Von Neumann equation $i\dot{\eta}_{tot}(t)=[\mathcal{H},\eta_{tot}(t)]$. However, in the weak coupling limit, the open Markovian dynamics of two UDW detector states $\eta_{ab}(t)$ can be extracted from $\eta_{tot}(t)$, by partial tracing over all environmental degrees of freedom (massless scalar field), that satisfies Kossakowski–Lindblad master equation \cite{gorini1976completely,lindblad1976generators}
\begin{equation}\label{mase}
	\frac{\partial \eta_{ab}(t)}{\partial t}=-i\left[ \mathcal{H}_{eff} , \eta_{ab}(t)\right] +\mathcal{L}\left[\eta_{ab}(t)\right],
\end{equation}
with 
\begin{equation}
\mathcal{H}_{eff}=\mathcal{H}_D- \frac{i }{2} \sum_{m, n = 1}^2 \Omega^{(mn)}_{ij} \hat{s}^{(m)}_i \hat{s}^{(n)}_j,   
\end{equation}
and
\begin{equation}
	\mathcal{L}\left[\eta_{ab}\right]=\sum_{\substack{i,j=1,2,3 \\ k,l=a,b}} \frac{\Omega_{ij}^{(mn)}}{2}\left[ 2\hat{s}_j^{(k)}\eta_{ab}\hat{s}_i^{(l)}-\left\lbrace \hat{s}_i^{(k)}\hat{s}_j^{(l)},\eta_{ab} \right\rbrace \right],
\end{equation}
represents the non-unitary evolution term produced by the coupling with external fields. The Kossakowski matrices $\Omega_{ij}^{(mn)}$ can be determined by the Fourier Transform of the following Wightman functions of the scalar field
\begin{equation}
Y^{(mn)}(t-t')=\left\langle 0\right| \chi(t,x^{(m)}) \chi(t',x^{(n)})\left|0 \right\rangle,
\end{equation}
and its Fourier Transform is given by
\begin{equation}
	\mathcal{Y}^{(mn)}(\omega)=\int_{-\infty}^{+\infty}d\Delta te^{i\omega\Delta t} Y^{(mn)}(\Delta t),
\end{equation}
here, the superscript $m,n=\left\lbrace a,b \right\rbrace $ labeling distinct detectors. For two-detector system, one can easily find that $Y^{(aa)}=Y^{(bb)}$ and $Y^{(ab)}=Y^{(ba)}$, which lead to $\mathcal{Y}^{(aa)}=\mathcal{Y}^{(bb)}\equiv\mathcal{Y}_{0}$ and $\mathcal{Y}^{(ab)}=\mathcal{Y}^{(ba)}$.

The master equation (\ref{mase}) enables us to describe the asymptotic equilibrium states of detectors at large times, which are governed by the competition between environment dissipation in CST background and QC generated through the Markovian dynamics of detectors \cite{benatti2004entanglement, benatti2003environment}. For a two-detector system, the initial interatomic separation $L\equiv\left|\textbf{x}^{(a)},\textbf{x}^{(b)} \right| $
is a control parameter of correlation generation, as Kossakowski matrices now become distance-dependent since, in general,  $\mathcal{Y}^{(ab)}=\mathcal{Y}^{(ba)}\equiv\mathcal{Y}(\omega, L)=\mathcal{Y}_{0}(\omega)f(\omega, L)$
for two separated detectors \cite{yu2011open, hu2013quantum}, where $f(\omega,L)$ is an even function of frequency $\omega$. It is not surprising \cite{benatti2005controlling} that the correlation generation between detectors would be more effective for smaller $L$, and becomes impossible for an infinitely large separation. It was shown in \cite{benatti2010entangling} that a proper $L$ always exists below which the generated correlation can persist asymptotically in final equilibrium states under environment dissipation. Therefore, we can concisely fix a small interatomic separation and only be concerned about the influence of environment decoherence on the equilibrium states of detectors. In such a situation, all the Kossakowski matrices become equal $\Omega_{ij}^{(aa)}=\Omega_{ij}^{(bb)}=\Omega_{ij}^{(ab)}=\Omega_{ij}^{(ba)}$ \cite{hu2011entanglement}, where
\begin{equation}
	\Omega_{ij}=\kappa_+\delta_{ij}-i\kappa_- \epsilon_{ijk}\delta_{3,k}+\tau\delta_{3,i}\delta_{3,j},
\end{equation}
with
\begin{equation}\label{gam}
	\kappa_{\pm}=\frac{1}{2}\left( \mathcal{Y}_{0}(\omega)\pm \mathcal{Y}(-\omega)\right) , \quad \tau=\mathcal{Y}_{0}(0)-\kappa_+.
\end{equation}
After solving the master equation (\ref{mase}), the final reduced density matrix of two detectors at asymptotic equilibrium can be expressed in the Bloch form \cite{benatti2004entanglement, benatti2003environment}

\begin{eqnarray}\label{e10}
\eta_{ab}(t)=\frac{1}{4}[\hat{s}_{0}^{(a)}\otimes\hat{s}_{0}^{(b)}+\sum_{j=1}^{3}\rho_{j}\Gamma_{j}+\sum_{i,j=1}^{3}\rho_{ij}\hat{s}_{i}^{(a)}\otimes\hat{s}_{j}^{(b)}],
\end{eqnarray}
where
\begin{eqnarray}\label{eq11}
	\rho_{j}=-\frac{R}{3+T^2}(\tau+3)\delta_{3,j},
\end{eqnarray}
and 
\begin{eqnarray}\label{eq12}
\rho_{ij}=\frac{1}{3+T^2}[T^2(\tau+3)\delta_{3,i}\delta_{3,j}+(\tau-T^2)\delta_{ij}],
\end{eqnarray}
Here, the ratio $T=\frac{\kappa_-}{\kappa_+}$ is determined by the dynamics of the system. The final equilibrium state also depends on the choice of initial state $\tau=\sum_{ii}\eta_{ii}(0)$, which is a constant of motion and satisfies $\tau\in [-3,1]$ 
to keep $\eta_{ab}(0)$ positive.

\subsection{QFI}
Here, we review the concept of QFI and its calculation based on the spectral decomposition of the density operator. We define QFI using a probe state represented by $\eta_\nu$, where $\nu$ is an unobservable parameter. The QFI concerning $\nu$ can be formulated as \cite{paris2009quantum}
\begin{equation}
\mathfrak{F}_\nu=Tr[L_\nu^2\eta_\nu],
\end{equation}
where the symmetric logarithmic derivative, denoted by \( L_\nu \), is defined through the equation \( \partial_\nu \eta_\nu = \frac{1}{2}(L_\nu \eta_\nu + \eta_\nu L_\nu) \). By expressing the density operator as \( \eta_\nu = \sum_{i} \mu_i \left| \psi_i \right\rangle \left\langle \psi_i \right| \), where \( \mu_i \) are the eigenvalues and \( \left| \psi_i \right\rangle \) are the corresponding eigenstates, an analytical solution for the QFI was obtained as shown in \cite{zhang2013quantum}.
\begin{equation}\label{fi}
\mathfrak{F}_\nu=\sum_{i}\frac{(\mu_i')^2}{\mu_i}+\sum_{i}\mu_i\mathfrak{F}_{\nu,i}-\sum_{i\neq j} \frac{8\mu_i\mu_j}{\mu_i+\mu_j}\left|\left\langle\psi_i'| \psi_j\right\rangle   \right|^2,
\end{equation}
with
\begin{equation}
\mu_i'=\partial_\nu\mu_i, \, \left| \psi_i'\right\rangle= \partial_\nu\left|\psi_i\right\rangle, \, \mathfrak{F}_{\nu,i}= 4(\left\langle \psi_i'| \psi_j\right\rangle-\left| \left\langle\psi_i'| \psi_i\right\rangle\right| ^2 ).
\end{equation}
It is evident that $\mathfrak{F}_\nu$ is solely dependent on the support set of $\eta_\nu$ and remains unaffected by eigenstates outside this support set. The first term in Eq. (\ref{fi}) depends only on the eigenvalues of $\eta_\nu$ and corresponds to the classical part, while the second and third terms correspond to the quantum part. Given the eigenstates described in (\ref{vpxs}), all the vectors $\left| \psi_i'\right\rangle=0$. Therefore, the QFI is entirely driven by the classical contribution 
\begin{equation}\label{fi}
\mathfrak{F}_\nu=\sum_{i}\frac{(\mu_i')^2}{\mu_i}.
\end{equation}

\subsection{LQU}
As a key measure in our analysis, we introduce LQU as the minimum skew information obtainable from a single local measurement. Let \(\eta_{ab}\) be the state of a bipartite system. Define \(K_{\Delta} = K_{\Delta}^{(a)} \otimes \hat{s}_{0}^{(b)}\) as a local observable, where \(K_{\Delta}^{(a)}\) is a Hermitian operator on the subsystem \((a)\) with spectrum \(\Delta\), and \(\hat{s}_{0}^{(b)}\) is the identity operator acting on subsystem \((b)\). We assume that \(\Delta\) is non-degenerate, corresponding to maximally informative observables on subsystem \((a)\). The LQU with respect to subsystem \((a)\), optimized over all local observables on \((a)\) with non-degenerate spectrum \(\Delta\), is expressed as \cite{Luo22003, Girolami2013}.

\begin{equation}
\mathfrak{L}(\eta_{ab}) \equiv \min_{ K_{\Delta}} \Upsilon(\eta_{ab},  K_{\Delta}), \label{LQU}
\end{equation}
where the skew information is defined as \cite{Luo22003, Wigner}
\begin{equation}
\Upsilon(\eta_{ab}, K_{\Delta})=-\frac{1}{2}{\rm
Tr}([\sqrt{\eta_{ab}}, K_{\Delta}]^{2}),
\end{equation}
Based on this Ref-\cite{Wigner}, we summarize the key characteristics of skew information: it is nonnegative, equals zero if and only if the state and the observable commute, and is convex, meaning it decreases under classical mixing. If a local operator \( K_{\Delta} \) exists such that \( \Upsilon(\eta_{ab}, K_{\Delta}) = 0 \), then the quantum system described by the state \( \eta_{ab} \) shows no QC between its two subsystems. For bipartite states, the LQU is expressed as \cite{Girolami2013}
\begin{equation} \label{eqs0}
  \mathfrak{L}\left( {{\eta_{ab}}}  \right) = 1 - \max \left( {{\vartheta_{11}},{\vartheta_{22}},{\vartheta _{33}}} \right).
  \end{equation}
where $\vartheta_{i,i=1,2,3}$ are the eigenvalues of the $3\times3$ symmetric matrix denoted $\Pi$, whose entries are defined as follows
\begin{equation}\label{w-elements}
 \Big(\Pi\Big)_{ij} \equiv  {\rm Tr} \left\{\sqrt{\eta_{ab}} \left(\hat{s}_{i}^{(a)}\otimes \hat{s}_{0}^{(b)}\right)\sqrt{\eta_{ab}} \left(\hat{s}_{j}^{(a)}\otimes
 \hat{s}_{0}^{(b)}\right)\right\},
\end{equation}
The Pauli matrices acting on subsystem \((a)\) are denoted by \(\hat{s}_{i}^{(a)} (i=x,y,z)\). 

\section{QFI and LQU in non-BD vacua \label{sec2}}
The dS-ST represents the simplest solution of Einstein's field equations when a positive cosmological constant, $\Lambda$, is present. dS-ST is maximally symmetric, characterized by the isometry group $\mathrm{SO}(4,1)$ in four dimensions. Its importance arises from two primary aspects. Firstly, the observed accelerated expansion of our Universe strongly suggests that it is influenced by a small, positive $\Lambda$ or an equivalent form of dark energy. Secondly, the remarkable uniformity and isotropy of the Universe on large scales, as observed in the CMB, imply that the early Universe likely underwent a phase of rapid accelerated expansion, commonly referred to as cosmic inflation, for which we suggest the reader to see the Ref-\cite{bhattacharya2020dirac} for further details.

\begin{figure}[H]
    \centering
    \includegraphics[width=0.99\linewidth]{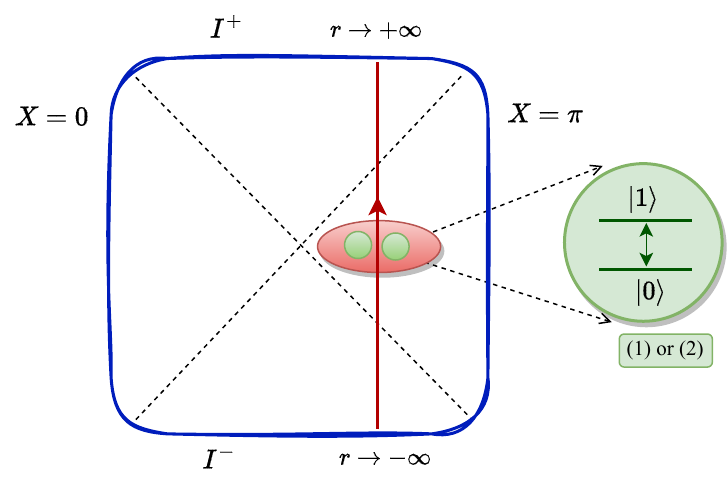}
 \caption{Schematic illustration of the Penrose diagram for the global dS-ST is presented in the coordinates (\ref{cord}), where the two UDW detectors, labeled (1) and (2), follow geodesic trajectories in dS-ST with a small interatomic separation ($L \ll l$). The infinities, \( I^{\pm} \), are space-like.}
    \label{fig:udd}
\end{figure}

Our benchmark system consists of freely falling UDW detectors in dS-ST, which interact weakly with a coupled massless scalar field. We employ the global coordinate system \((r, X, \theta, \phi)\), where the detectors are co-moving with the expansion. The line element of dS-ST is given by
\begin{equation}\label{cord}
	ds^2=dr^2-\frac{\cosh^2(Hr)}{H^{2}}\left[dX^2+\sin^{2}X(d\theta^{2}+\sin^{2}\theta d\phi^{2}) \right],
\end{equation}
where the Hubble parameter \( H \) defines a positive cosmological constant \( \Lambda = 3H^2 \) and a radius of curvature  \( l = H^{-1} \). If the initial separation between detectors is much smaller than the radius of curvature, i.e., $ L \ll l $, the resulting correlation at asymptotic equilibrium becomes independent of $ L $. As a result, the detectors' final equilibrium states take the form given in Eqs. (\ref{e10}) and (\ref{eq11}-\ref{eq12}) (see Fig. \ref{fig:udd}).
The scalar wave equation can then be resolved in coordinates (\ref{cord}), defining a dS-invariant BDV $\left| BD\right\rangle $. In the massless, conformal coupling limit, for a freely falling detector, the associated Wightman function $Y^+_{BD}(x(t),y(t'))\equiv \left\langle BD\right|\chi(x(t))\chi(y(t'))\left| BD\right\rangle$ becomes \cite{allen1987massless}
\begin{equation}\label{e13}
	Y^+_{BD}(x(t),y(t'))=-\frac{H^2}{16\pi^2\left[ (t-t')H/2-i\epsilon\right] },
\end{equation}
which fulfills Kubo-Martin-Schwinger
(KMS) condition, i.e., $Y^+_{BD}(t')\equiv Y^+_{BD}(t'+i\beta)$, implying that for a detector in BDV sector, it ends up in thermal equilibrium at universal GH temperature $T_{GH}\equiv1/\beta=1/2\pi l$ \cite{gibbons1977cosmological}.

By Mottola–Allen transformation, one can further define a one-parameter family of de Sitter-invariant vacua called $\alpha-$vacua $\left| \alpha\right\rangle $, each of which can be interpreted as a squeezed state over $\left| BD\right\rangle $, i.e., $\left| \alpha\right\rangle=\hat{S}(\alpha)\left| BD\right\rangle $, where $\text{Re}(\alpha) < 0$ and $\hat{S}(\alpha)$ denotes a squeezing operator in quantum optics \cite{danielsson2002consistency,einhorn2003interacting, einhorn2003squeezed,goldstein2003note,collins2004taming}. Throughout this work, we adopt CPT invariant $\alpha-$vacua, which means $\alpha<0$ is real, with $|\alpha|$ denoting the magnitude of $\alpha$ in all figures and analyses. Therefore, the Wightman function for the scalar field in $\alpha-$vacua can be expressed in terms of $Y^+_{BD}$ as

\begin{equation}\label{sh}
	\begin{array}{cc}
		\begin{aligned}
			Y^+_{\alpha}(x,y)& =N^{-1}\left[ Y^+_{BD}(x,y)+e^{2\alpha} Y^+_{BD}(y,x) \right.\\
			&\left. e^{\alpha}\left(Y^+_{BD}(x,y_A)+ Y^+_{BD}(x_A,y) \right) \right] 
		\end{aligned}
	\end{array},
\end{equation}
where subscript $\alpha$ denotes a particular choice of $\alpha$ value, $N\equiv 1-e^{2\alpha} $ 
and $x_A$ is the antipodal point of $x$. As $\alpha\rightarrow -\infty$, we find that $Y^+_{\alpha}(x,y)$ reduces to the Wightman function in BDV $Y^+_{BD}(x,y)$, which uniquely extrapolates to the same short-distance behavior of two-point correlation function in the Minkowski vacuum, as the curvature of de Sitter vanishing. Employing the relation \cite{bousso2002conformal} 
\begin{equation}
	Y^+(x,y_A)=Y^+(x_A,y)=Y^+(t-i\pi),
\end{equation}
and substituting (\ref{e13}) into (\ref{sh}), we can calculate the Fourier transformation of the Wightman function in non-BD sectors as
\begin{equation}
	\mathcal{Y}=\frac{\omega(1+e^{\alpha-\pi \omega})^{2}}{2\pi (1-e^{-\beta\omega})(1-e^{2\alpha})}.
\end{equation}
The related Kossakowski coefficients are
\begin{equation}\label{she}
	\begin{array}{cc}
		\begin{aligned}
			\Sigma_{+,\alpha}& =\frac{\omega\left[ (1+e^{\alpha-\pi \omega})^{2}+e^{-\beta\omega}(1+e^{\alpha+\pi \omega})^2\right] }{4\pi (1-e^{-\beta\omega})(1-e^{2\alpha})},\\
			\Sigma_{-,\alpha}& =\frac{\omega\left[ (1+e^{\alpha-\pi \omega})^{2}-e^{-\beta\omega}(1+e^{\alpha+\pi \omega})^2\right] }{4\pi (1-e^{-\beta\omega})(1-e^{2\alpha})},\\
			T_{\alpha}& =\frac{ (1+e^{\alpha-\pi \omega})^{2}-e^{-\beta\omega}(1+e^{\alpha+\pi \omega})^2 }{(1+e^{\alpha-\pi \omega})^{2}+e^{-\beta\omega}(1+e^{\alpha+\pi \omega})^2 }.
		\end{aligned}
	\end{array}
\end{equation}
For fixed energy spacing of detectors, it is easy to find that $T_{\alpha}\in \left[ -1,+1\right] $ as $\beta$ varies. Inserting (\ref{she}) into (\ref{e10}), we obtain the final equilibrium state of two detectors respecting to $\alpha-$vacua, which is
\begin{equation}\label{xs}
	\eta_{ab}(t)=\left(
	\begin{array}{cccc}
		\eta_{-} & 0 & 0 & 0 \\
		0 & \eta_{22} &\eta_{23}  & 0 \\
		0 & \eta_{23} & \eta_{22} & 0 \\
		0 & 0 & 0 & \eta_{+}
	\end{array}
	\right),
\end{equation}
\begin{equation}
\eta_{\pm}=\frac{(3+\tau)(T_{\alpha}\pm1)^2}{4(3+T_{\alpha}^2)},\quad
	 \eta_{23}=\frac{\tau-T_{\alpha}^2}{2(3+T_{\alpha}^2)},
\end{equation}
\begin{equation}
\eta_{22}=\frac{3-\tau-(1+\tau)T_{\alpha}^2}{4(3+T_{\alpha}^2)}.
\end{equation}
The eigenvalues of the  final state (\ref{xs}) are given by
\begin{equation}\label{eid}
\begin{split}
    &\mu_1= \frac{(1-T_{\alpha})^2 (\tau +3)}{4 \left(T_{\alpha}^2+3\right)},\quad \mu_2= \frac{(T_{\alpha}+1)^2 (\tau +3)}{4 \left(T_{\alpha}^2+3\right)},\\
    & \mu_3= \frac{(1-T_{\alpha}^2) (\tau +3)}{T_{\alpha}^2+3}, \quad \quad \mu_4= \frac{1-\tau }{4}.
\end{split}
\end{equation}
while the corresponding eigenvectors in the computational basics $\{ |0\rangle, |1\rangle\}$ are
\begin{equation} \label{vpxs}
\begin{split}
    &|\psi\rangle_1= |00\rangle,\quad \quad |\psi\rangle_2= |11\rangle,\\
    & |\psi\rangle_3= \frac{1}{\sqrt{2}}(|10\rangle-|01\rangle), \quad |\psi\rangle_4=\frac{1}{\sqrt{2}}(|01\rangle-|10\rangle).
\end{split}
\end{equation}
Before proceeding to the QFI and LQU analysis of our model, it is necessary to consider several critical observations regarding $\alpha$-vacua. First, the squeezing nature of $\alpha$-vacua can significantly constrain measurement uncertainty, suggesting that certain intrinsic QCs may be concealed in these non-Bell diagonal states \cite{maldacena2013entanglement,kanno2014entanglement}. Second, $\alpha$-vacua does not exhibit thermal characteristics, as evidenced by the failure of (\ref{sh}) to satisfy the KMS condition unless it reduces to the BDV. This deviation from thermality has led to various attempts \cite{martin2001trans,kaloper2003initial,goldstein2003initial,ashoorioon2014non} to consider $\alpha$-vacua as an alternative initial state for inflation. To account for the anticipated correction in the primordial power spectrum of order $\sim \mathcal{O}(H/\Lambda)^2$, the parameter $\alpha$ can be directly linked to $\Lambda$, which represents some fundamental scales of new physics (e.g., the Planck scale or the string scale) \cite{danielsson2002inflation,danielsson2002note}.

We choose to estimate the GH parameter $\beta$ for any arbitrary initial state. The related $\mathfrak{F}_\beta$ can be straightforwardly calculated from (\ref{fi}), by substituting the eigenvalues (\ref{eid}) of the diagonalized density matrix of the detectors system (\ref{xs}). According to different preparations of the initial state encoded in $\tau \in [-3,1]$, we come to two classes of QFI: \\
(i) $\tau = -3$, only one non-vanishing eigenvalue $\mu_4 = 1$, which gives $\mathfrak{F}_\beta= 0$; \\
(ii) \(\tau \in (-3, 1]\), all eigenvalues (\ref{eid}) are non-vanishing. 
Since \(\partial_\beta \mu_4 = 0\), we have

\begin{equation}\label{qfb}
\mathfrak{F}_\beta=\sum_{i=1,2,3}\frac{(\mu_i')^2}{\mu_i}=\frac{2 (\tau +3) \left(R_\alpha^2-3\right) \partial_ \beta R_\alpha^2}{\left(R_\alpha^2-1\right) \left(R_\alpha^2+3\right)^2}.
\end{equation}

The explicit form of the matrix elements (\ref{w-elements}), which depend on the system parameters from (\ref{xs}) and the GH decoherence environment, is derived as 
 \begin{align}
\vartheta_{11}&=\vartheta_{22}=\frac{1}{4} \left(\sqrt{ \eta_{-}}+\sqrt{ \eta_{+}}\right) \notag \\
	&\quad \times\left(\sqrt{1-\tau }+\sqrt{\frac{\left(1-R_\alpha^2\right)
   (\tau +3)}{R_\alpha^2+3}}\right),
	\label{eq7}
\end{align}
and 
\begin{equation}\label{w22}
\vartheta_{33}=\frac{\tau +3}{2}-\frac{ (\tau +3)}{R_\alpha^2+3}+\sqrt{\frac{(1-\tau)\left(1-R_\alpha^2\right) (\tau +3)}{4(R_\alpha^2+3)}},
\end{equation}
while $\vartheta_{ij}=0$ for $i\neq j$.

For $\tau \in [-3,1]$, the situation for (\ref{qfb}) and (\ref{eq7}--\ref{w22}) is complicated due to its dependence on the infinite family of non-BDV. In the following, our aim to demonstrate that for certain initial states, QCs can be generated in the final state (\ref{xs}) after the Markovian evolution of the system. To illustrate this, we consider two UDW detectors initially prepared in a Bell-diagonal state on a freely falling basis
\begin{equation}\label{bes}
\eta_{ab}(0) = \frac{1}{4} \left( \hat{s}_{0}^{(a)} \otimes \hat{s}_{0}^{(b)} + \sum_{i=1}^3 r_i \, \hat{s}_{i}^{(a)} \otimes \hat{s}_{i}^{(b)} \right),
\end{equation}
with coefficients \( 0 \leq |r_i| \leq 1 \). This state is the convex combination of four Bell states and reduces to maximally entangled states (Bell-basis) if \( |r_1| = |r_2| = |r_3| = 1 \). 
\section{Results and analysis \label{sec3}}
In this section, we investigate the impact of GH decoherence on the QFI and LQU of scalar fields in dS-ST. To achieve this, we examined freely falling UDW detectors in dS-ST, which interact weakly with a massless scalar field in the dS-invariant BDV. This analysis allowed us to evaluate the degradation of QFI and LQU due to the GH decoherence as a function of various physical parameters, including GH temperature $T_{GH}=1/\beta$, the initial state preparation $\tau$, $\alpha$-vacua, and the detector energy spacing $\omega$.

\subsection{QFI}
First, we evaluate how the QFI depends on various parameters to assess its effectiveness as a diagnostic tool for understanding the dS-ST structure. To illustrate our results, we will provide plots of the QFI versus the inverse of temperature $\beta$ and different relevant system parameters. 

\begin{widetext}
\begin{minipage}{\linewidth}
\begin{figure}[H]
\centering
\subfigure[]{\label{figure4a}\includegraphics[scale=0.45]{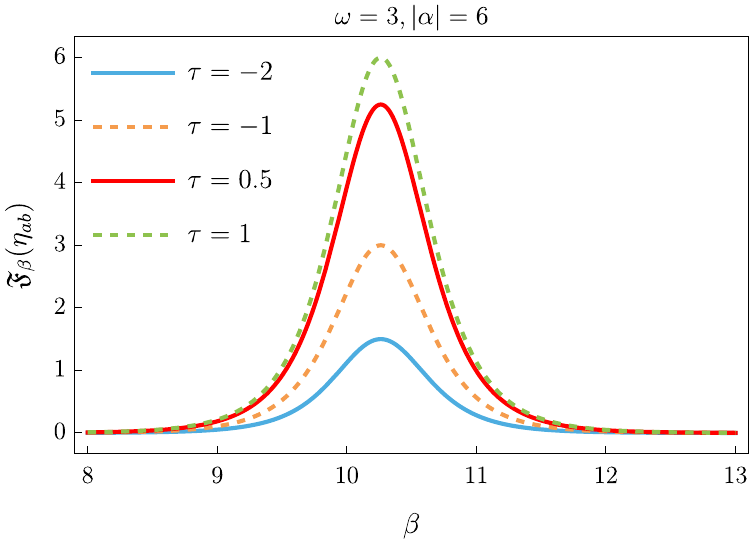}}
\subfigure[]{\label{figure4b}\includegraphics[scale=0.45]{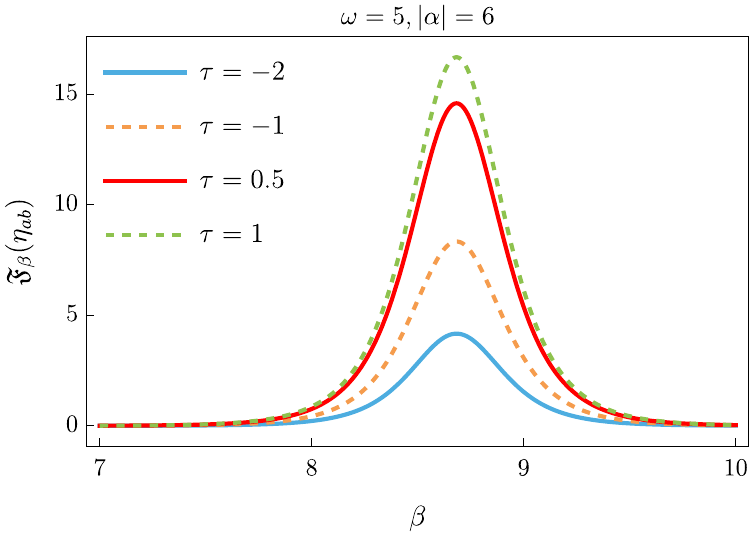}}
\subfigure[]{\label{figure4c}\includegraphics[scale=0.45]{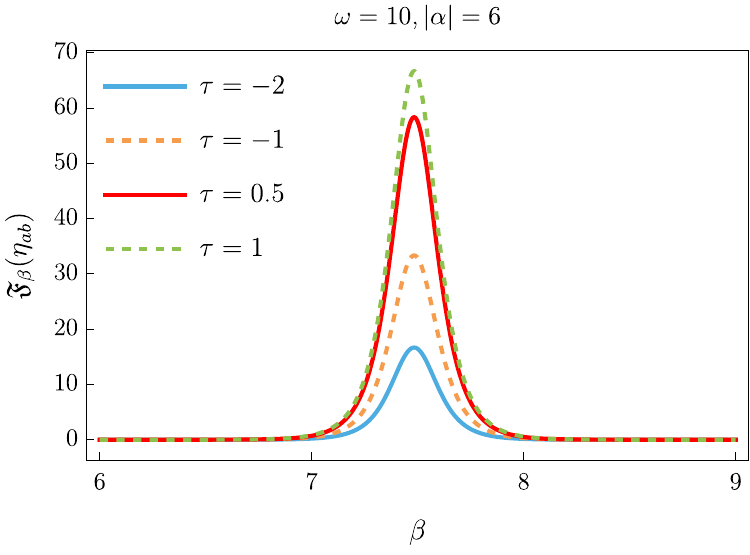}}
\caption{The plot of QFI among UDW detectors as a function of $\beta$ for various fixed values of $\tau$, namely $\tau = -2$ (solid blue), $\tau = -1$ (dashed orange), $\tau = 0.5$ (solid red), and $\tau = 1$ (dashed green). In the left panel \ref{figure4a}, we set $\omega = 3$ and $|\alpha| = 6$; in the middle panel \ref{figure4b}, $\omega = 5$ and $|\alpha| = 6$; and in the right panel \ref{figure4c}, $\omega = 10$ and $|\alpha| = 6$.}
\label{figure4}
\end{figure}
\end{minipage}
\end{widetext}
To gain deeper insights into the influence of the GH temperature and the choice of the initial state \(\tau\) on the QFI, we present in Fig. \ref{figure4} the behavior of the QFI as a function of the parameter \(\beta\) and the initial state selection \(\tau\), while keeping the non-BDV sectors fixed at \(|\alpha| = 6\). The QFI values are computed for fixed detector energy spacings, specifically \(\omega = 3, 5, 10\). For all examined cases, a local peak in the QFI emerges as \(\beta\) decreases, indicating that the optimal precision for estimating \(\beta = 2\pi l\) is attained when the detectors are positioned at a smaller curvature radius \(l\). Furthermore, the QFI exhibits a global maximum at an intermediate value, denoted as \(\beta_{\text{int}}\), beyond which it asymptotically decreases. Notably, for \(\tau < 0\) (e.g., \(\tau = -2, -1\)), the QFI peaks become more pronounced, rising sharply before undergoing a rapid decline near \(\beta_{\text{int}}\). Additionally, the position of \(\beta_{\text{int}}\) is observed to shift with increasing detector energy gap \(\omega\). This shift modifies the location of the QFI peak while simultaneously enhancing the QFI magnitude, ultimately leading to a decrease in \(\beta_{\text{int}}\) within non-BDV sectors. These findings indicate that the optimal estimation of \(\beta\) is highly sensitive to the choice of the initial state. Moreover, as depicted in Fig. \ref{figure4a}, the QFI initially increases to a maximum before decreasing as \(\beta\) grows, while it continues to rise with \(\tau\). This trend suggests that higher precision in estimating \(\beta\) can be achieved through an appropriate selection of \(\tau\). In conclusion, the analysis clearly demonstrates that the QFI is significantly influenced by both the initial state of the UDW detector and the GH temperature.
\begin{widetext}
	\begin{minipage}{\linewidth}
		\begin{figure}[H]
			\centering
		\subfigure[]{\label{figure5a}\includegraphics[scale=0.55]{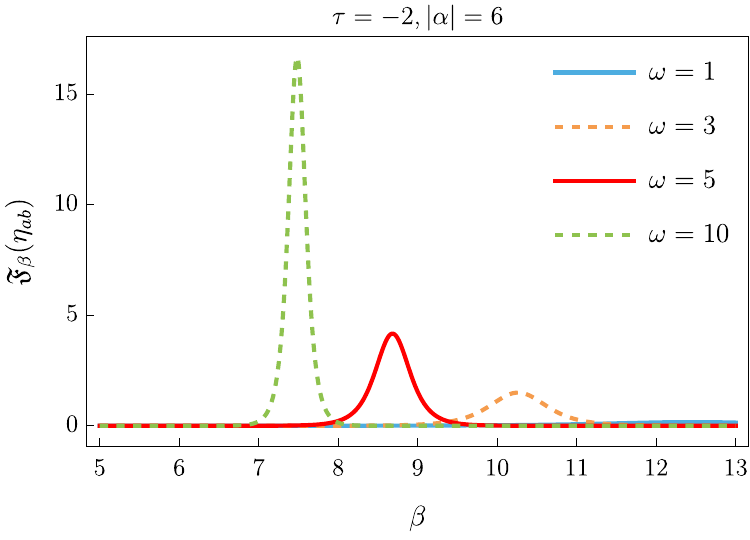}} \quad \quad \quad
		\subfigure[]{\label{figure5c}\includegraphics[scale=0.55]{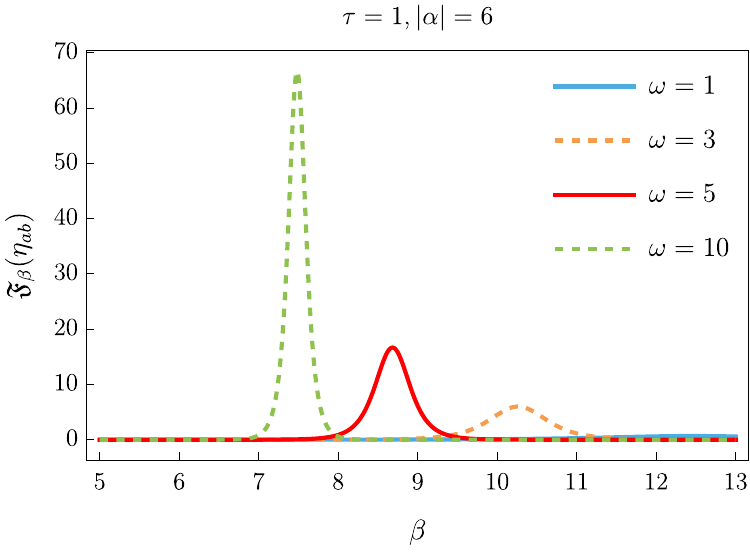}}\\
        \subfigure[]{\label{figure6a}\includegraphics[scale=0.55]{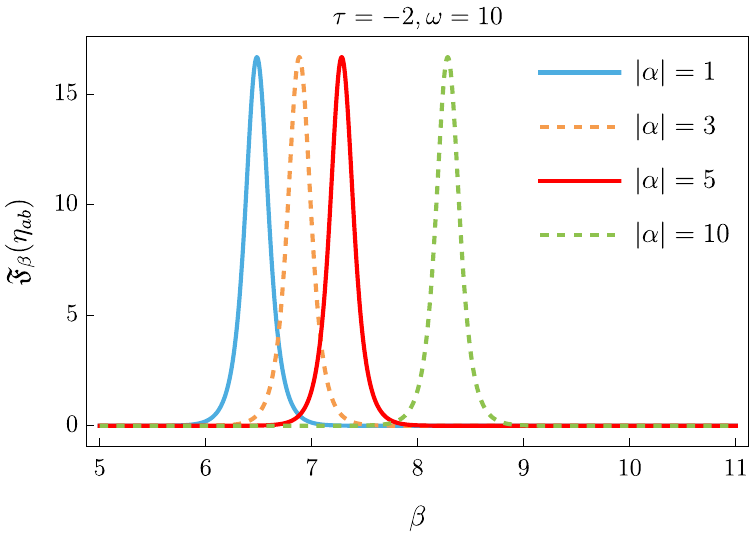}}\qquad\qquad
		\subfigure[]{\label{figure6c}\includegraphics[scale=0.55]{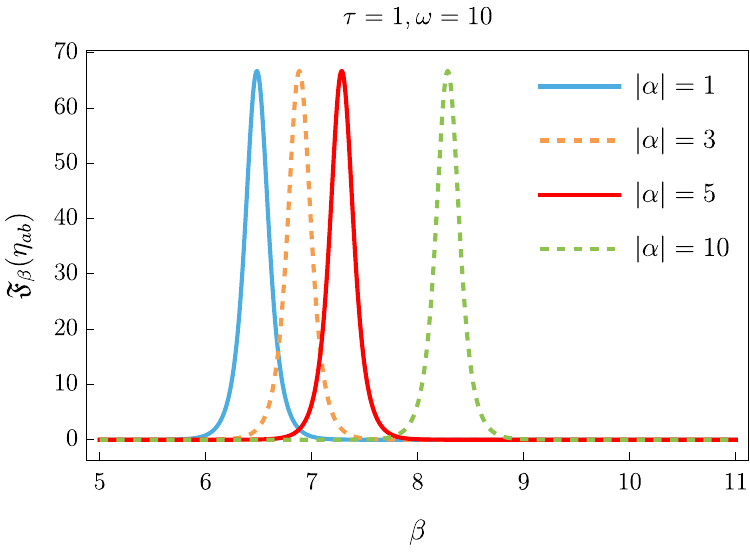}}
\caption{The plots of QFI among UDW detectors as a function of $\beta$ for various fixed parameters. Panels \ref{figure5a} and \ref{figure5c} depict QFI for different fixed values of $\omega$, namely $\omega = 1$ (solid blue), $\omega = 3$ (dashed orange), $\omega = 5$ (solid red), and $\omega = 10$ (dashed green). In panel \ref{figure5a}, we set $\tau = -2$ and $|\alpha| = 6$, while in panel \ref{figure5c}, $\tau = 1$ and $|\alpha| = 6$. Panels \ref{figure6a} and \ref{figure6c} show QFI for different fixed values of $|\alpha|$, namely $|\alpha| = 1$ (solid blue), $|\alpha| = 3$ (dashed orange), $|\alpha| = 5$ (solid red), and $|\alpha| = 10$ (dashed green). In panel \ref{figure6a}, we set $\tau = -2$ and $\omega = 10$, while in panel \ref{figure6c}, $\tau = 1$ and $\omega = 10$.}
\label{f5}
\end{figure}
\end{minipage}
\end{widetext}
To elucidate the relationship between the energy spacing \(\omega\), different choices of superselection sectors of \(\alpha\)-vacua, and the QFI of a pair of UDW detectors in dS-ST, we present the QFI of \(\eta_{ab}\) as a function of \(\omega\) and \(\alpha\) in Fig. \ref{f5}. This analysis is conducted for two distinct cases: (i) when \(\tau=-2\), and (ii) when \(\tau=1\). Fig. \ref{f5} illustrates the estimation precision of the GH decoherence parameter, \(\beta\), as quantified by the QFI for various values of \(\omega\) and \(|\alpha|\). As depicted in Fig. \ref{figure5a}, an increase in the energy spacing \(\omega\) enhances the estimation accuracy of \(\beta\). Furthermore, a larger \(\omega\) shifts the peak QFI towards lower values of \(\beta\), implying that higher precision in parameter estimation can be achieved for a detector with a greater GH temperature. This result is consistent with the trends observed in Fig. \ref{figure4}, where increasing \(\tau=1\) leads to a substantial rise in the maximum QFI. The influence of \(\alpha\), which corresponds to different superselection sectors of dS vacua, on the QFI for \(\beta\) is further examined in Figs. \ref{figure6a} and \ref{figure6c} for \(\tau=-2\) and \(\tau=1\), respectively. The QFI exhibits a single peak that shifts towards lower values of \(\beta\) as \(|\alpha|\) increases. Notably, for \(\tau=-2\), the maximum QFI remains approximately constant at \(\sim 15\), whereas for \(\tau=1\), it undergoes a significant increase, reaching values around \(\sim 70\).  

As illustrated in Figs. \ref{figure4} and \ref{f5}, the behavior of the QFI exhibits a distinct Gaussian trend. A detailed analysis of the dependence of QFI on various system parameters reveals that increasing the value of \(\tau\) enhances the maximum QFI. This suggests that QFI is fundamentally influenced by the temperature \(\beta\), the energy spacing \(\omega\), and different choices of superselection sectors of \(\alpha\)-vacua. Notably, a larger \(\tau\) enhances the system's quantum information processing capabilities.  Furthermore, increasing the parameter \(|\alpha|\) shifts the peak of the QFI towards higher values of the GH temperature, whereas an increase in \(\omega\) causes the peak to shift towards lower GH temperatures. This implies that both the structure of the \(\alpha\)-vacua and the detector’s energy spacing modulate the probability distribution of quantum system interactions, thereby contributing to the enhancement of QFI.

\subsection{LQU}
Next, we display that the LQU arising from $\alpha$-vacua can be harnessed as a resource in RQI tasks. We expect that the amount of LQU will exhibit dependence on the initial state preparation of the detectors and the various superselection sectors of $\alpha$-vacua.

\begin{widetext}
	\begin{minipage}{\linewidth}
		\begin{figure}[H]
			\centering
			\subfigure[]{\label{figure1a}\includegraphics[scale=0.45]{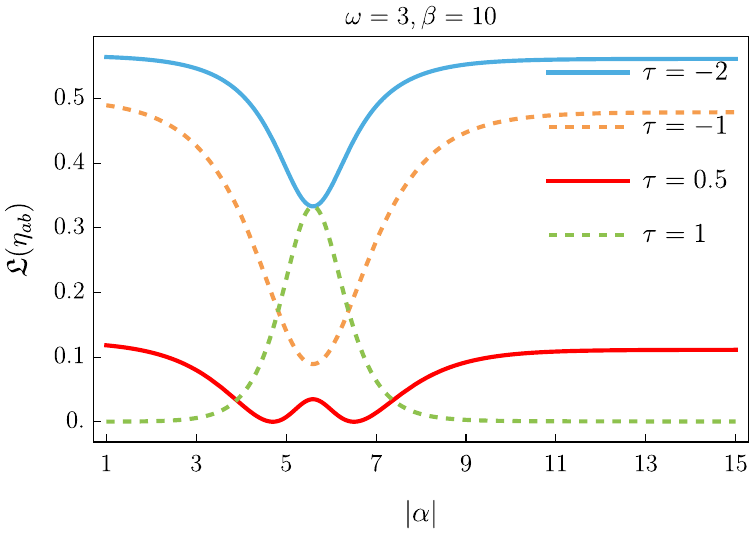}}
			\subfigure[]{\label{figure1b}\includegraphics[scale=0.45]{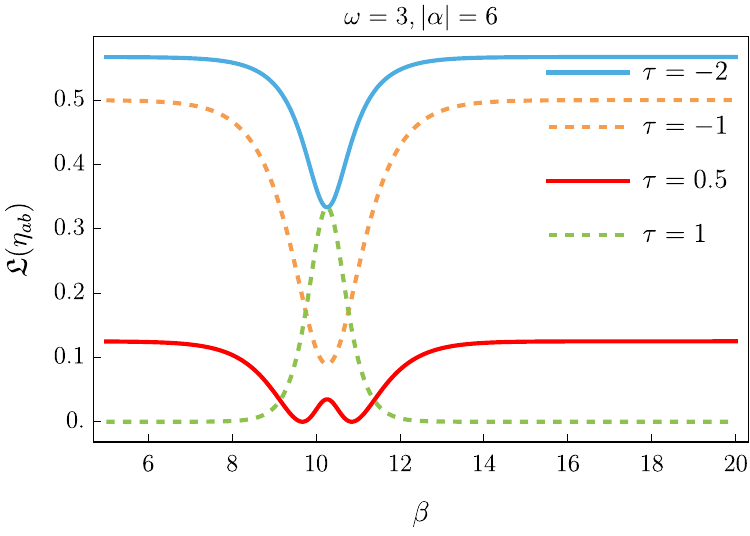}}
		\caption{The plots of LQU as a function of $|\alpha|$ (panel \ref{figure1a}) and $\beta$ (panel \ref{figure1b}) for various fixed values of the parameter $\tau$, namely $\tau = -2$ (solid blue), $\tau = -1$ (dashed orange), $\tau = 0.5$ (solid red), and $\tau = 1$ (dashed green). In both panels, $\omega = 3$, with $\beta = 10$ in \ref{figure1a} and $|\alpha| = 6$ in \ref{figure1b}.}
\label{f1}
\end{figure}
\end{minipage}
\end{widetext}
To gain a deeper understanding of the dependence of QCs between two uniformly accelerated UDW detectors in dS-ST, we analyze the LQU for different initial state choices, denoted by \(\tau\), while maintaining a fixed energy spacing of \(\omega = 3\). Fig. \ref{figure1a} presents the variation of LQU as a function of \(|\alpha|\) for a fixed decoherence parameter \(\beta = 10\). The results indicate that for negative values of \(\tau\), the LQU is higher compared to positive \(\tau\). In the negative \(\tau\) regime, LQU initially decreases with increasing \(|\alpha|\). However, beyond \(|\alpha| > 5.5\), LQU starts to increase again, eventually reaching its maximum limit and stabilizing at different values depending on the initial state. On the other hand, for \(\tau = 1\), the LQU exhibits a single peak within the interval \(|\alpha| \in [4,8]\). Outside this range, LQU vanishes, signifying a transition of the quantum system to a classical state. Similarly, Fig. \ref{figure1b} illustrates the behavior of LQU as a function of the GH decoherence parameter \(\beta\), with a fixed value of \(|\alpha| = 6\). The qualitative behavior remains consistent with that observed in Fig. \ref{figure1a}, except that the peak of LQU shifts to the interval \(\beta \in [8,12]\). These observations suggest that transitioning from negative to positive values of \(\tau\) leads to a reduction in LQU, which implies the onset of decoherence and a shift toward classical behavior. Furthermore, the influence of \(|\alpha|\)-vacua and the GH decoherence parameter \(\beta\) appears to affect both the generation of LQU and its maximum attainable value in a similar manner. Notably, the decoherence environment exhibits both constructive and destructive effects on LQU. Specifically, for negative initial states, LQU increases with the GH effect. This result challenges the conventional belief that environmental decoherence solely leads to a reduction in LQU in dS-ST.  
\begin{widetext}
	\begin{minipage}{\linewidth}
		\begin{figure}[H]
			\centering
			\subfigure[]{\label{figure2a}\includegraphics[scale=0.45]{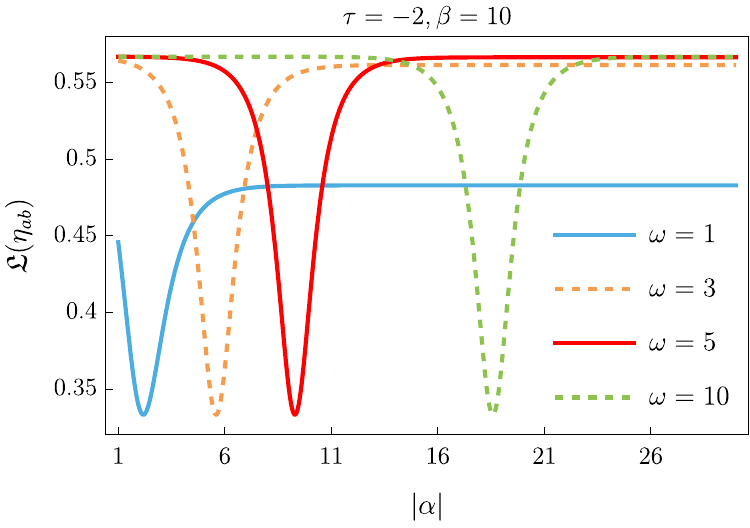}}
			\subfigure[]{\label{figure2b}\includegraphics[scale=0.45]{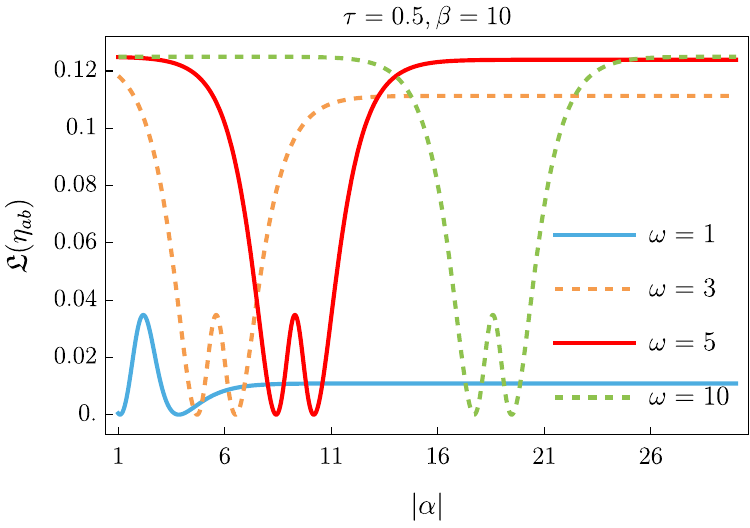}}
			\subfigure[]{\label{figure2c}\includegraphics[scale=0.45]{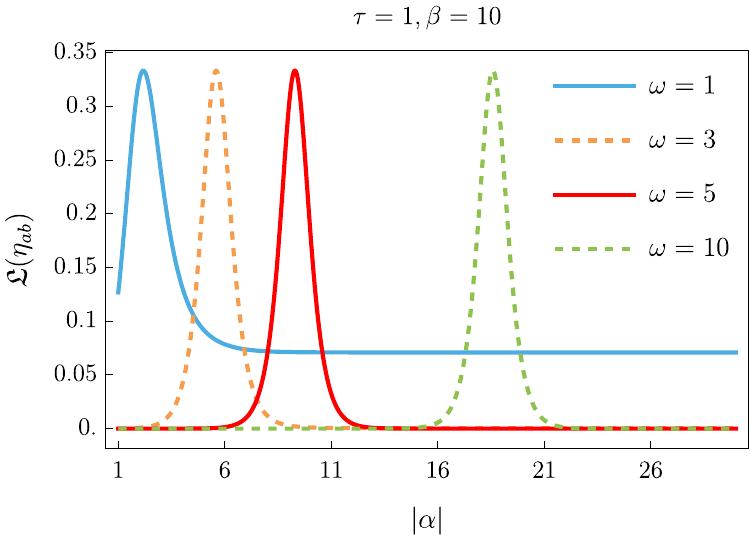}}
		\caption{The plot of LQU among UDW detectors as a function of $|\alpha|$ for various fixed values of $\omega$, namely $\omega = 1$ (solid blue), $\omega = 3$ (dashed orange), $\omega = 5$ (solid red), and $\omega = 10$ (dashed green). In the left panel \ref{figure2a}, we set $\tau = -2$ and $\beta = 10$; in the middle panel \ref{figure2b}, $\tau = 0.5$ and $\beta = 10$; and in the right panel \ref{figure2c}, $\tau = 1$ and $\beta = 10$.}         \label{f2}
		\end{figure}
	\end{minipage}
\end{widetext}
The impact of varying values of $\omega$ and $\tau$ on LQU as a function of $|\alpha|$, with $\beta = 10$, is illustrated in Fig. \ref{f2}. For $\tau = -2$ and different values of $\omega$, Fig. \ref{figure2a} shows that lower $\omega$ values result in reduced LQU. However, as the energy spacing increases, the maximum LQU rises, peaking at $0.57$. Additionally, there is a transient dip that shifts toward higher $|\alpha|$ values as $\omega$ increases. Conversely, Fig. \ref{figure2b} indicates that when $\tau = 0.5$, the maximum LQU values decrease. For lower $\omega$ values, the LQU approaches zero, which indicates that the system is nearing a classical state and experiencing increased decoherence. In contrast, Fig. \ref{figure2c} reveals that for $\tau = 1$, lower $\omega$ values yield higher LQU compared to higher $\omega$ values. As $\omega$ increases, there is still a shift in LQU toward higher $|\alpha|$ vacuum values. These results lead us to conclude that by jointly adjusting the energy spacing $\omega$ and the $|\alpha|$ vacuum for a specific choice of the initial state $\tau$, we can significantly enhance the LQU of the system. This approach offers a viable solution to reduce the negative effects of the GH temperature on the two-detector system in CST.
\begin{widetext}
	\begin{minipage}{\linewidth}
 		\begin{figure}[H]
			\centering
			\subfigure[]{\label{figure3a}\includegraphics[scale=0.45]{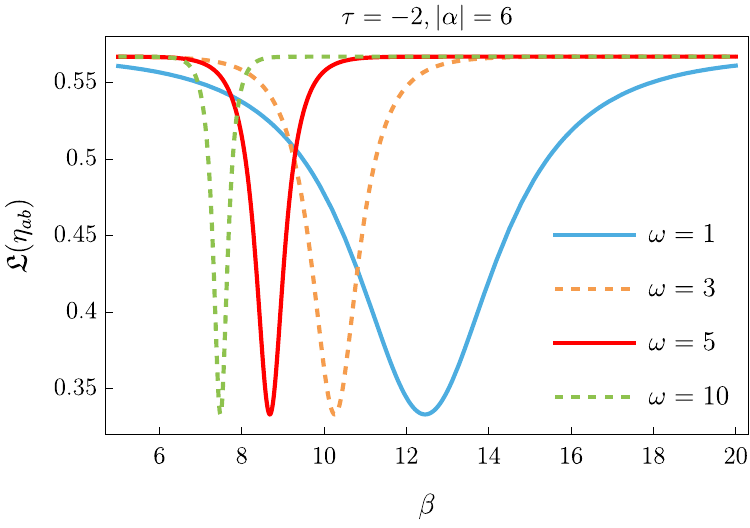}}
			\subfigure[]{\label{figure3b}\includegraphics[scale=0.45]{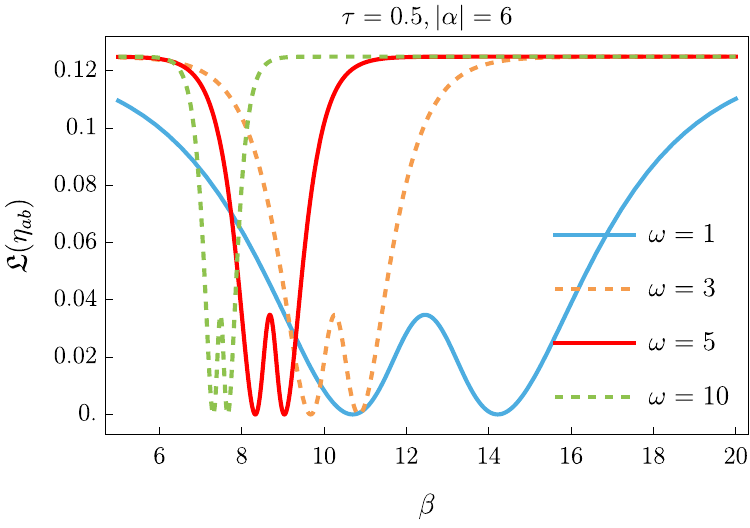}}
			\subfigure[]{\label{figure3c}\includegraphics[scale=0.45]{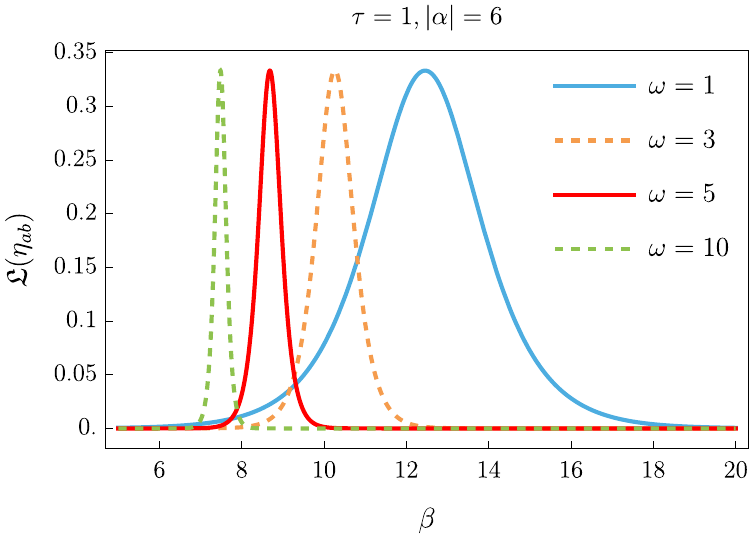}}
		\caption{The plot of LQU among UDW detectors as a function of $\beta$ for various fixed values of $\omega$, namely $\omega = 1$ (solid blue), $\omega = 3$ (dashed orange), $\omega = 5$ (solid red), and $\omega = 10$ (dashed green). In the left panel \ref{figure3a}, we set $\tau = -2$ and $|\alpha| = 6$; in the middle panel \ref{figure3b}, $\tau = 0.5$ and $|\alpha| = 6$; and in the right panel \ref{figure3c}, $\tau = 1$ and $|\alpha| = 6$.}			\label{f3}
		\end{figure}
	\end{minipage}
\end{widetext}

Now, we examine the impact of varying the values of \( \omega \) and \( \tau \) on the LQU as a function of the GH decoherence \( \beta \) with \( |\alpha| = 6 \). Figure \ref{f3} shows that, generally, an increase in \( \omega \) enhances the LQU of the system, thus reducing decoherence. Specifically, Fig. \ref{figure3a} reveals a pit-like structure when \( \tau = -2 \); the width of this pit diminishes as the energy spacing increases and the bip shifts to lower \( \beta \) values. Conversely, positive values of \( \tau \) influence the inversion of LQU. Figure \ref{figure3b} demonstrates that the bip begins to broaden, and a small peak starts to form. However, when \( \tau = 1 \), Fig. \ref{figure3c} shows that the LQU is inverted, where the minimum bounds of LQU become the maximum bounds as \( \tau \) increases. From this figure, it is evident that LQU represents a dynamic system significantly influenced by the control parameter \( \tau \). Continuous adjustments to \( \tau \) induce topological transformations in the system's LQU. Initially, the function exhibits a concave (pit-like) profile, indicative of a stable equilibrium point resembling a narrow well for large \( \omega \) values. As \( \tau \) transitions from negative to positive values, this well broadens, forming a small peak at the curve's center. This peak gradually intensifies until the function assumes a convex (peak-like) shape at \( \tau = 1 \). In this state, the peak represents the quantum region, while the remaining portion of the curve signifies the classical nature of the system.

\section{Conclusion}\label{sec4}
In this paper, we investigated the thermal nature of the GH effect using QFI and LQU to evaluate the intrinsic estimation and correlations of non-BDV in dS-ST. We considered accelerating UDW detectors as open quantum systems coupled to massless scalar fields in the de Sitter-invariant vacuum, with the complete dynamics resolved from a Lindblad master equation for the density matrix. 

As a result, we have observed that the QFI of detectors can achieve the same maximum for different choices of de Sitter-invariant vacuum sectors, given fixed values of $\tau$ and $\omega$. Moreover, the peak maximum of QFI can be enhanced by selecting appropriate energy spacing. However, GH decoherence generally reduces the QFI, which converges to an asymptotic value that is independent of the detector’s initial state preparation. From a metrological perspective, this indicates that enhanced precision in estimating the GH effect can be achieved with relatively high energy spacing. Finally, we emphasize that when the initial state preparation is positive, specifically for $\tau = 1$, the related QFI is robust against GH decoherence in the sense that its local peak can persist over a very long range. This robustness of QFI can significantly facilitate practical relativistic quantum estimation tasks.
Similarly, in our LQU analysis, we found that LQU exhibits a striking revival phenomenon as the GH temperature increases. In particular, the amount of LQU can drastically improve after a certain turning point, $\beta_c$, for non-BD sectors due to the competition between GH decoherence and the intrinsic correlation of $\alpha$-vacua. For specific choices of the detectors' initial state, the measure of LQU will first decrease to zero, then increase, and eventually approach an asymptotic value. Notably, LQU in specific non-BD sectors can persist even under arbitrarily large GH decoherence.

It was pointed out that the qualitative effects of the dS horizon and its associated thermality have been previously reported in the context of black hole spacetimes, particularly in Schwarzschild-de Sitter black holes \cite{bhattacharya2022entanglement}. Specifically, they suggest that the interplay of multiple horizons (black hole and cosmological) and their associated temperatures can lead to novel entanglement dynamics that deviate from standard asymptotically flat or anti-dS black hole cases. We aim to revisit these issues in future works by investigating the robustness of QFI and LQU under multiple horizons thermality, along with their implications for RQM.

\section*{Disclosures}
The authors declare that they have no known competing financial interests.

\section*{Data availability}
No datasets were generated or analyzed during the current study.

\section*{Acknowledgments}
M.Y. Abd-Rabbou was supported in part by the University of Chinese Academy of Sciences. 

\bibliography{references}
\bibliographystyle{unsrt}

\end{document}